\documentclass[aps,prl,twocolumn,nofootinbib]{revtex4-2}
\usepackage[utf8]{inputenc}
\usepackage{amsmath}
\usepackage{amsfonts}
\usepackage{amssymb}

\usepackage{environ}
\usepackage{tikz}
\tikzset{Witten diagram/.style={execute at begin picture={%
\draw[blue,fill=white!20] circle[radius=\pgfkeysvalueof{/tikz/Witten/radius}];
}},vertex/.style={circle,fill,inner sep=1.5pt,node
contents={}},
Witten/.cd,radius/.initial=1cm}
\NewEnviron{wittendiagram}[1][]{\vcenter{\hbox{\begin{tikzpicture}[Witten diagram,#1]%
\BODY
\end{tikzpicture}}}}

\begin{document}

\title{Bulk renormalization and the AdS/CFT correspondence    }

\author{Máximo Bañados${}^{*}$, Ernesto Bianchi${}^{* \dagger }$, Iván Muñoz${}^{* \dagger }$ and Kostas Skenderis${}^{\dagger }$}
\affiliation{$*$ Facultad de Física, Pontificia Universidad Católica de Chile, Santiago, Chile}  
\affiliation{$\dagger$ STAG Research Centre \& Mathematical Sciences, $\mbox{Highfield}$, University of Southampton, SO17 1BJ Southampton, UK}
\affiliation{maxbanados@gmail.com; \ E.Bianchi@soton.ac.uk; \ iimunoz1@uc.cl;\ K.Skenderis@soton.ac.uk}

\begin{abstract}
We develop a systematic renormalization procedure for QFT in anti-de Sitter spacetime. UV infinities are regulated using a geodesic point-splitting method, which respects AdS isometries, while IR infinities are regulated  by cutting off the radial direction (as in holographic renormalization). The renormalized theory is defined by introducing $Z$ factors for all parameters in the Lagrangian and the boundary conditions of bulk fields (sources of dual operators), and a boundary counterterm action, $S_{\rm ct}$, such that the limit of removing the UV and IR regulators exists. The results are in general scheme dependent (mirroring the analogous result in flat space) and require renormalization conditions. These may be provided by the dual CFT (or by string theory in AdS).  Our analysis amounts also to a first principles derivation of the Feynman rules regarding Witten diagrams. The presence and treatment of IR divergences is essential for correctly accounting for anomalous dimensions of dual operators. We apply the method to scalar $\Phi^4$ theory and obtain the renormalized 2-point function of the dual operator to 2-loops, and the renormalized 4-point function to 1-loop order, for operators of any dimension $\Delta$ and bulk spacetime dimension up to $d+1=7$.

\end{abstract}

\maketitle

The AdS/CFT correspondence \cite{JM1,Gubser:1998bc, Witten, AdSCFTBible} relates string theory on $(d+1)$-dimensional anti-de Sitter (AdS) spacetime (times a compact space) and conformal field theory (CFT)  in $d$ dimensions. At low energies this becomes a relation between AdS gravity and strongly coupled CFT.  The holographic dictionary links parameters $p^i$ in the bulk action (masses and couplings) with CFT data $C_j$ (conformal dimensions and OPE coefficients) and 
the fields $\varphi_0$ parametrising the boundary conditions of bulk fields $\Phi$ with sources of boundary gauge invariant operators ${\cal O}$.  Then the bulk partition function $Z[\varphi_0; p^i]$ is identified with the generating functional of CFT correlation functions,
\begin{equation}
Z[\varphi_0;p^i] = \left \langle \exp\left(-\int_{\partial AdS} \varphi_0 {\cal O} \right)\right \rangle_{C_J}\, .
\end{equation}
Here the r.h.s. is the path integral over the CFT at the conformal boundary of AdS, specified by the CFT data $C_J$. The CFTs that enter the AdS/CFT correspondence typically admit a large $N$ 't Hooft limit, and bulk loops correspond to $1/N^2$ corrections.

This relation however needs renormalization and its precise form has only been fully developed at tree level in the bulk/leading large $N$ limit in the boundary CFT \cite{deHaro:2000vlm,Kostas}.  There has been continuous progress about AdS/CFT at loop level in recent years, mostly based on new results regarding CFT correlators at subleading order in the large $N$ limit \cite{Penedones:2010ue,Fitzpatrick:2011dm, Aharony:2016dwx, Alday:2017xua, Aprile:2017bgs, Aprile:2017xsp, Giombi:2017hpr, Yuan:2017vgp, Yuan:2018qva, Bertan:2018khc,  Bertan:2018afl, Carmi:2018qzm, Ghosh:2018bgd, Ponomarev:2019ofr, Carmi:2019ocp, Meltzer:2019nbs, Albayrak:2020bso, Meltzer:2020qbr, Costantino:2020vdu, Carmi:2021dsn, Fichet:2021pbn,Fichet:2021xfn, Heckelbacher:2022fbx, Frob:2022onc}, but there has been no systematic discussion of the bulk side. The purpose of this paper is to provide such a systematic discussion. Our discussion follows (on purpose) as close as possible textbook discussions of renormalizability in flat space, but as we will see there are important new issues.

A successful setup not only makes possible a meaningful application of the duality, where both sides are well defined, but also provides structural support to the duality, independent of the specifics of any particular example.  An important property of holographic dualities is the so-called UV/IR connection \cite{Susskind:1998dq}: UV infinities of one theory are linked to IR of the other and vice versa. An essential property of local Quantum Field Theory (QFT) is that the UV infinities are local and an important general result of holographic renormalization is that at tree level in the bulk (and for arbitrary $n$-point functions) bulk IR infinities due to the infinite volume of spacetime are local \cite{deHaro:2000vlm, Papadimitriou:2004ap}.  When considering the duality at loop order there are a number of similar structural relations that need to be satisfied.

On the bulk side, there are now potentially both UV and IR divergences. The IR divergences corresponding to boundary UV divergences should continue to be local. UV divergences in the bulk correspond to IR divergences in the boundary QFT, and in QFT one does not add counterterms for such divergences: they should cancel on their own. The bulk theory should thus be UV finite, suggesting that in the full duality the bulk should have a string theoretic description.  At low energies however, where the bulk is described by a supergravity theory there are UV infinities at loop order and one would like to understand how to treat them. 

In a CFT two- and three-point functions are completely fixed by conformal invariance, up to constants, and higher points are fixed up to functions of cross ratios. One would thus expect to be able to obtain the same results in the bulk just using bulk isometries and we will see that this is indeed the case. Conformal invariance is broken by conformal anomalies and these are accounted by holographic renormalization \cite{Henningson:1998gx}. Renormalization of bulk UV infinities however should respect conformal symmetry, and indeed we will see that there is a bulk UV regulator that respects all AdS isometries. 

The CFT data, the dimensions of operators and the constants and functions of cross ratios that appear in the correlation functions may  receive $1/N^2$ corrections, and our purpose is to discuss how these renormalize from loops in the bulk. We will use the $\Phi^4$ theory in a fixed AdS background \footnote{One may formally decouple dynamical gravity by taking the limit of the Planck mass going to infinity (or equivalently Newton's constant to zero, $G_N \to 0$) keeping fixed (and independent of $G_N$) the parameters that enter in the Lagrangian of the matter fields (as in (\ref{regS}) below). With these conventions, matter propagators and gravity-matter vertices are independent of $G_N$, vertices involving only gravitons scale as $G_{N}^{-1}$ and the graviton propagator as $G_N$, and one may check that diagrams with internal gravitons are suppressed. The AdS isometries then imply that  correlators in a fixed AdS background satisfy the conformal Ward identities on their own. The application of our method to perturbative gravity is technically more involved but it can be done along the same lines and it will be presented elsewhere. In particular, the geodesic point-splitting method we discuss below also regulates graviton loops.} to illustrate the method but the methodology applies generally. We will find that  this specific theory is renormalizable to 1-loop order in bulk spacetime dimensions up to seven in the sense that all UV infinities that appear in the computation of boundary correlators up to 4-point functions can be removed by local bulk counterterms. One in general needs to renormalize both the bulk parameters $p^i$, the masses and the couplings that appear in the bulk action, and  the sources $\varphi_0$.

\paragraph{Regulators:} It is essential that we introduce both a UV and an IR regulator. The IR regulator is the usual holographic radial cut-off. Using the AdS metric, $d s^2 = \ell^2 (dz^2 + d \vec{x}^2)/z^2$, we restrict the radial integration to $z \geq \varepsilon$.
For the UV cut-off we modify the bulk-to-bulk propagator by displacing one of the points along a geodesic with affine parameter $\tau$.
Let $x_1, x_2$ be two points in AdS and $u(x_1, x_2) =  \left((z_1-z_2)^2 + (\vec{x}_1-\vec{x}_2)^2\right)/2 z_1 z_2$ the AdS invariant distance. Consider now the geodesic,
\begin{equation} \label{geod}
z(\tau)=\frac{z}{\cosh(\tau/\ell)}, \qquad \vec{x}(\tau)=\vec{x}+z\tanh(\tau/\ell)\ \hat{n}\, ,
\end{equation}
where $\hat{n}$ is a unit vector that is orthogonal to $(\vec{x}_1{-}\vec{x}_2)$, $\hat{n}{\cdot}(\vec{x}_1-\vec{x}_2)=0$, and $\ell$ is the AdS radius that we set to one from now on. A direct computation yields
$ u(x_1(\tau),x_2)=-1+\cosh \tau \left[1+u(x_1,x_2)\right]$.
Note that $u(x_1(\tau),x_2)=0$ iff $x_1=x_2$ and $\tau=0$, so $\tau$ acts as a short-distance AdS invariant regulator. In terms of the chordal distance, $\xi(x_1, x_2)=1/(1+u(x_1,x_2))$, $\xi(x_1(\tau), x_2)=\xi(x_1, x_2)/\cosh \tau $. Loop diagrams are made out of bulk-to-bulk propagators $G(x_1, x_2)$ and bulk vertices. The UV regulator only affects the bulk-to-bulk propagators, which now become 
\begin{equation}
G_\tau(x_1, x_2){\equiv}G(x_1(\tau), x_2){=}G\left(\frac{\xi(x_1, x_2)}{\cosh \tau}\right){\equiv}G_\tau(\xi)\, ,
\end{equation}
where $x_1(\tau)$ is given in (\ref{geod}) and $G(x_1, x_2)$ is the standard bulk-to-bulk propagator,
$G(\xi) =  2^{-\Delta} c_\Delta \xi^\Delta/(2\Delta -d)\, _2F_1\left( {\Delta/2}, {(\Delta +1)/2}, \Delta -{d/2}+ 1 , \xi^2\right)$ and $c_\Delta = \Gamma (\Delta )/(\pi ^{d/2} \Gamma (\Delta -d/2 ))$.

\paragraph{Renormalization:} We consider $\Phi^4$ theory with action,
\begin{align}\label{regS}
    &S[\Phi] = \int  d^{d+1} x \sqrt{g}
    \left[\frac{1}{2}\partial_\mu\Phi \partial^\mu\Phi
    +\frac{m_0^2}{2}\Phi^2+\frac{\lambda_0}{4!}\Phi^4\right]\, .
\end{align}
The regulated bulk partition function is given by
\begin{align}
    &Z^{\rm reg}_\text{AdS}\left[\varphi_0;m_0^2,\lambda_0; \varepsilon, \tau\right] = \nonumber \\
    &\int_{\Phi\sim\phi[\varphi_0]}\! \! \! \! \! \! \! \! \! \! \! \!
    D\Phi\ \exp{\left(- 
    S[\Phi] - \int_{z=\varepsilon} \! \! \! \! \! \! \! d^d x {\cal L}_{ct}[\Phi]\right)}  \, ,\label{Z_reg}
\end{align}
where $\phi[\varphi_0]$ specify the boundary conditions (to be discussed below) and ${\cal L}_{\rm ct}$ are boundary counterterms. These were originally introduced in the process of holographic renormalization \cite{deHaro:2000vlm, Kostas}, but (more fundamentally) are required for the variational problem to be well posed \cite{Papadimitriou:2005ii, Andrade:2006pg}.  To renormalize the theory we treat the parameters that enter the theory, $\varphi_0, m_0^2, \lambda_0$ as bare parameters, and define the renormalized parameters   $\varphi, m, \lambda$ via,
\begin{equation}\label{CT1}
    \varphi_0
    =Z_\varphi \varphi, \
    \ m_0^2 
    =  m^2+\delta m^2, \ 
    \lambda_0 
      =\lambda+\delta\lambda \, .
\end{equation}
$Z_\varphi, \delta m, \delta \lambda$  depend on regulators and $\varphi, m, \lambda$ are finite. We assume  $\lambda$  is small and work perturbative to order $O(\lambda^2)$. At tree level $\delta m=\delta \lambda=0, Z_\varphi=1$.

The renormalized theory is now defined by
\begin{equation}
Z^{\rm ren}_\text{AdS}\left[\varphi;m^2,\lambda\right]=\lim_{\varepsilon \to 0} \lim_{\tau \to 0}
Z^{\rm reg}_\text{AdS}\left[\varphi_0;m_0^2,\lambda_0; \varepsilon, \tau\right]\, , 
\end{equation}
and  renormalized CFT correlators can be obtained by functionally differentiating $Z^{\rm ren}_\text{AdS}$ w.r.t. $\varphi(\vec{x})$.
The theory is renormalizable if we can carry out this program without introducing new terms in the bulk action. We will see that this is the case to 1-loop order for the $\Phi^4$ theory up to seven (bulk) dimensions. We emphasise that bulk subtractions lead to scheme dependence. To fix the  renormalized parameters and thus the scheme dependence we need renormalization conditions. These can be provided either by the full string theory in AdS or by the dual CFT: the renormalized mass is fixed by the spectrum of theory, or equivalently by the spectrum of dimensions of the dual CFT, $\varphi$ is fixed by the normalization of the 2-point function, and $\lambda$ can be related to the OPE coefficients  of the dual CFT. 

\paragraph{Perturbative computation:}

We split the field into its classical $\phi$ and quantum $h$ parts, 
\begin{equation}\label{classquan}
\Phi = \phi[\varphi_0] + h\, .
\end{equation}
The classical field $\phi[\varphi_0]$ solves the equations of motion
\begin{equation} \label{feq}
(-\Box + m^2)\phi = - \frac{\lambda}{3!} \phi^3 \, ,
\end{equation}  
with the non-normalizable boundary condition $\phi(z,\vec{x}) \sim z^{d-\Delta} \varphi_0(\vec{x})$ as $z\rightarrow 0$, with $\varphi_0(\vec{x})$ being the source for the dual operator.  The quantum fluctuation $h$ on the other hand satisfies normalizable boundary conditions, $h(z, \vec{x}) \sim z^\Delta \check{h}(\vec{x})$.

Using \eqref{classquan} the partition function \eqref{Z_reg} becomes,
\begin{align}\label{Z3}
   & Z^{\rm reg}_\text{AdS}
    = e^{-S_{\rm sub}[\phi; \varepsilon, \tau]}  \int Dh \exp\biggl[-S[h]  \\
&-\int_x \left(\delta m^2\phi+\frac{\delta\lambda}{6}\phi^3\right)h 
   +(\lambda + \delta \lambda) \left(\frac{1}{6}\phi h^3+\frac{1}{4}\phi^2h^2\right)\biggr]\, , \nonumber
\end{align}
where $\int_x = \int_{z  \geq \varepsilon}\! \! d^{d+1} x \sqrt{g}$, $S[h]$ is the action (\ref{regS}) with $\Phi$ replaced by $h$, $S_{\rm sub} = S^{\rm on-shell}_{\rm reg} + S_{\rm ct}$ is the subtracted on-shell action (as in  \cite{Bianchi:2001kw}), and $S^{\rm on-shell}_{\rm reg}$ is the regulated on-shell action. This term gives the tree-level Witten diagrams, see \cite{Kostas}. Here our focus is on computing the loop contributions. 

We are interested in computing quantum corrections to CFT correlators, up to 4-point functions to order $\lambda^2$. As the CFT source is the leading term in the classical field $\phi$, it suffices to expand \eqref{Z3} to 4th order in $\phi$ and then integrate out $h$. The path integral over $h$ is a straightforward application of QFT methods and expresses the answer in terms of bulk correlation functions. 
While  the bulk-to-bulk propagator $G(x_1,x_2)$ is divergent at coincident points
the regulated propagator, $G_\tau(x,x) = G_\tau(1)$ is finite, and this suffices to regulate all UV infinities.  Moreover,
 the regulated propagator is invariant under transformations that transforms simultaneously $x_1$ and $x_2$ by AdS isometries, so 
 bulk loop diagrams do not break any of the AdS isometries.  The regulation of coincident points by replacing $\xi$ by $\xi/(1+\epsilon)$ in the bulk-to-bulk propagator was introduced in  \cite{Sachs}, and our analysis relates this regulator  to geodesic point splitting.

To proceed we add a source term $\int J \, h$ to the action and following the usual QFT manipulations we arrive at the expression, 
\begin{eqnarray}\label{W4}
&&e^{W_{\rm reg}} = e^{-S_{\rm sub}} e^{-\int_x \left(\delta m^2\phi+\frac{\delta\lambda}{6}\phi^3\right) {\delta \over \delta J} + \frac{\delta m^2}{2}{\delta^2 \over \delta J^2}} \\ 
&&
e^{-\int_x \frac{\lambda{+}\delta \lambda}{4} \left(\phi^2 + \frac{2}{3}\phi {\delta \over \delta J} + {1 \over 6}{\delta^2 \over \delta J^2}\right){\delta^2 \over \delta J^2}} \left. e^{{1 \over 2} \int_{x} \int_{y} J(x) G_\tau(x,y) J(y) }   
 \right|_{J=0} \, . \nonumber
\end{eqnarray}
Here the source $J(x)$ is only an intermediate device. The true source is the boundary function $\varphi_0$ inside the classical field $\phi[\varphi_0]$. To a given order $\lambda^p$ one has to compute several functional derivatives with respect to $J$.  Once all derivatives have been computed, $J$ is set to zero and one is left with a (non-local) functional of the classical field $\phi$. Keeping the terms relevant for the computation of 2- and 4-point function through order $\lambda^2$ we get
\begin{eqnarray}\label{Wl2}
&&W_{\rm reg} = -S_{\rm sub}
- \frac{1}{2}\left(\delta m^2+\frac{\lambda {+} \delta \lambda}{2} G_\tau(1)\right)  \int_{x}\phi^2(x) \nonumber \\
&& +\frac{1}{2} \left(\delta m^2+\frac{\lambda}{2} G_\tau(1)\right) 
\int_{x_1}  \int_{x_2} \left(\frac{\lambda}{2}  \phi(x_1)^2 G^2_\tau(x_1, x_2)  
\right. \nonumber \\
&&\left. +\phi(x_1) G_\tau(x_1, x_2)  \phi(x_2)  \left(\delta m^2+\frac{\lambda}{2} G_\tau(1)\right)\right)    \nonumber   \\
&& +\frac{\lambda^2}{12} \int_{x_1} \int _{x_2} \phi(x_1) G_\tau^3(x_1, x_2)  \phi(x_2) -\frac{\delta \lambda}{4!} \int_{x_1}  \phi(x_1)^4 \nonumber \\
&& + \frac{\lambda^2}{16} \int_{x_1} \int_{x_2}  \phi(x_1)^2 G_\tau^2(x_1, x_2)  \phi(x_2)^2 + \ {\cal O}(\lambda^3)\, , 
\end{eqnarray} 
where $G_\tau(1)=G_\tau(x,x)$ is the regulated value of the bulk-to-bulk propagator at coincident points. Note that this is not the final form of the perturbative series. The classical field $\phi$ is itself a series in $\lambda$, obtained solving perturbatively in $\lambda$ Eq. (\ref{feq}). 

\begin{figure*}[t]
\includegraphics[width=\textwidth]{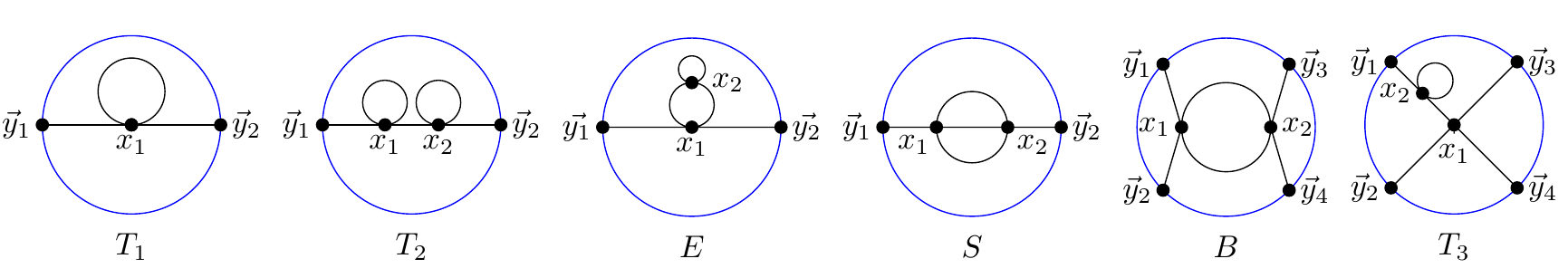}
  \caption{Witten diagrams contributing to the 2- and 4-point function to order $\lambda^2$. \label{fig1}}
  \end{figure*}

Using this result we may now compute the boundary correlators up  to 4-point functions to order $\lambda^2$. Differentiating w.r.t. sources we find that the boundary correlators are represented by the expected Witten diagrams with the same symmetry factors as Feynman diagrams
(internal lines joined by bulk-to-bulk propagators  and external lines joined by bulk-boundary propagator), a result which has now been derived from first principles.  All relevant diagrams are listed in Fig. \ref{fig1} and we shortly discuss their evaluation. Correlators of this type have been calculated in various works in the past \cite{Penedones:2010ue,Fitzpatrick:2011dm, Aharony:2016dwx, Alday:2017xua, Aprile:2017bgs, Aprile:2017xsp, Giombi:2017hpr, Yuan:2017vgp, Yuan:2018qva, Bertan:2018khc,  Bertan:2018afl, Carmi:2018qzm, Ghosh:2018bgd, Ponomarev:2019ofr, Carmi:2019ocp, Meltzer:2019nbs, Albayrak:2020bso, Meltzer:2020qbr, Costantino:2020vdu, Carmi:2021dsn, Fichet:2021pbn,Fichet:2021xfn,Heckelbacher:2022fbx, Frob:2022onc}, 
and while part of our methodology is present in many of these papers, no previous work contains a complete and coherent discussion of all issues. In particular, in most of the existing literature, the UV regulator is often {\it ad hoc}, only regularisation but not renormalisation was done, scheme dependence was not discussed, and the importance of IR regulator was overlooked. One of our main results  is that IR divergences are essentially responsible for the appearance of anomalous dimensions in correlators, as one may anticipate based on the fact that they correspond to boundary UV divergences.

We start with the 2-point function. The general form for this function, to all orders in $\lambda$, is 
\begin{eqnarray}\label{O2}
C_2(\vec{y}_1,\vec{y}_2) = \int_{x_1}\!  \int_{x_2}\! \! K(\vec{y}_1,x_1) K(\vec{y}_2,x_2)  P_2(x_1,x_2) 
\end{eqnarray}
where $K(\vec{y}_1,x_1)= c_\Delta z_1^\Delta/(z_1^2 + |\vec{x}_1-\vec{y}_1|^2)^\Delta$  is the bulk-to-boundary propagator and  $P_2(x_1,x_2)$ is the ``amputated'' bulk-to-bulk 2-point function.
$P_2(x_1,x_2)$ is an integral over all internal vertices of products of bulk-to-bulk propagators that  join the points $x_1$ and $x_2$ to themselves and the internal vertices.
As long as we use the regulated bulk-to-bulk propagator, this expression is UV finite.  
Recall that the $G_\tau(x_1, x_2)$ is invariant under AdS isometries, and so is $P_2(x_1, x_2)$.
 Assuming (\ref{O2}) is IR finite, one may extract the $\vec{y}_1$ and $\vec{y}_2$ dependence from the integral by 
simple manipulations: first shift the integration variables, $\vec{x}_i \to \vec{x}_i +\vec{y}_2, i=1, 2$, and then rescale, $x_i \to x_i |\vec{y}_{1}-\vec{y}_2|$, where both transformations are AdS isometries.   After these manipulation the integral no longer depends on the external points (but still depends on the UV regulator) and we will call its value $A_2(\tau)$.
Altogether we obtain,
\begin{equation}\label{irf}
C_2(\vec{y}_1,\vec{y}_2) =  A_2(\tau)  y_{12}^{-2\Delta} \, ,
\end{equation}
where $y_{12} = |\vec{y}_2 - \vec{y}_1|$.
One may now renormalize this correlator by just rescaling the source $\varphi_0$ ({\it i.e.} using $Z_\varphi$).
This is the expected form of the 2-point function of an operator of dimension $\Delta$. Here however $\Delta$ is the tree-level dimension, and we thus find that $\Delta$ does not renormalize to all orders: if there were no IR divergences, there would be no anomalous dimensions. 

This analysis is however not correct because (\ref{O2}) is IR divergent and a cutoff is needed. The transformations needed to arrive at (\ref{irf}) act on the integration limits and the naive invariance under AdS isometries is broken. At one-loop order, the relevant diagram is the tadpole diagram $T_1$ (see Fig. \ref{fig1}), and by explicit evaluation we find,
\begin{align} \label{tadpole}
&T_1 = \int_{z_1  \geq \varepsilon}\! \! d^{d+1} x_1 \sqrt{g} K(x_1,\vec{y}_1) G_\tau(x_1, x_1) K(x_1,\vec{y}_2) \nonumber \\
&=G_\tau(1) \left(\frac{\varepsilon^{-(2\Delta-d)}}{(2\Delta-d)}\delta(\vec{y}_1-\vec{y}_2)+\dots 
\right. \\
&\left.  -\frac{c_\Delta}{\lvert\vec{y}_1-\vec{y}_2\rvert^{2\Delta}}  \left[\ln\left(\frac{\varepsilon}{\lvert\vec{y}_1-\vec{y}_2\rvert}\right)^2+\psi(\Delta)-\psi(\nu)\right] + \cdots\right)\, , \nonumber 
\end{align}
where  $\nu=\Delta-d/2$.
It is no longer possible to remove the infinities by only rescaling  the source $\varphi_0$ and renormalization of the mass is now required. 
 
Similar manipulations, now involving also inversions, show that, barring IR divergences, 3-points functions \footnote{The 3-point function vanishes in the theory (\ref{regS}) because the action is invariant under $\Phi \to -\Phi$. In theories with non-vanishing 
3-point function (for example $\Phi^3$ theory) the argument  shows that the 3-point function would have the form dictated by conformal invariance.} and 4-point functions take the expected form, with $\Delta$ the tree-level dimension. Renormalization produces anomalous dimensions for $\Delta$ and corrections to the constants appearing in the 2- and 3-point functions and the function of cross ratios in higher point functions.

We are now ready to present the results of the evaluation of the Witten diagrams in Fig. \ref{fig1} for any $\Delta$ and $d$.
The theory is renormalizable up to $d+1=7$ bulk dimensions and the counterterms that remove the infinities are given by
\begin{widetext}
\begin{align}
\delta\lambda &= \frac{\lambda^2}{2} \left(3\text{Div}\left[a_0(\tau)\right]+\frac{\left(2\Delta-\frac{d}{2}\right)}{\Delta}\text{Div}\left[a_1(\tau)\right]\right)+\lambda^2G_1\, ;  \\
\delta m^2 = &-\frac{\lambda}{2}\text{Div}\left[G_\tau(1)\right]
 - \frac{\lambda^2}{4}\left( \text{Div}\left[\left(3\text{Div}\left[a_0(\tau)\right]+\frac{\left(2\Delta-\frac{d}{2}\right)}{\Delta}\text{Div}\left[a_1(\tau)\right]\right)G_\tau(1) \right] + 2 G_1 \text{Div}\left[G_\tau(1)\right]\right)
\nonumber\\
    &+\frac{\lambda^2}{4}\text{Div}\left[\text{Con}\left[G_\tau(1)\right]E(\tau)\right]+\frac{\lambda^2}{6}\text{Div}\left[S(\tau)\right]+\frac{\lambda^2}{2}F_1\text{Div}\left[E(\tau)\right]+\lambda F_1+\lambda^2F_2\, ,
\end{align}
\end{widetext}
where Div[$f(\tau)$] denotes the divergent part of the function $f(\tau)$ as $\tau \to 0$ and Con[$f(\tau)$] denotes the part that has a limit as $\tau \to 0$. Such a split is always ambiguous because one may add a finite piece to Div[$f(\tau)$] and subtract it from Con[$f(\tau)$]. This ambiguity  is encoded by the functions $G_1, F_1, F_2$ which represent scheme dependence. To fix these functions one needs renormalization conditions, as noted earlier.  The functions $a_0(\tau), a_1(\tau), E(\tau), S(\tau)$ are defined by
 \begin{align} \label{integrals}
    &\int_{x_2} G_\tau^2(x_1,x_2) = E(\tau)\, ;\\
    &\int_{x_2} G_\tau^3(x_1,x_2)K(x_2,\vec{y}_2) = S(\tau)K(x_1,\vec{y_2})\, ; \nonumber\\
    & \int_{x_2} G_\tau^2(x_1, x_2)K(x_2,\vec{y}_3)K(x_2,\vec{y}_4) = \chi K(x_1,\vec{y}_3)K(x_1,\vec{y}_4)\, , \nonumber
\end{align}
where $\chi = \sum_{i=0}^\infty (a_i(\tau)+b_i(\tau)\log \hat{\chi}) \hat{\chi}^i$, $\hat{\chi} = \tilde{K}(x_1,\vec{y}_3)\tilde{K}(x_1,\vec{y}_4)y_{34}^2$ and $\tilde{K}(x,\vec{y})= z/(z^2 + |\vec{x} -\vec{y}|^2)$. The result for the integrals in (\ref{integrals}) is fixed by AdS isometries, {\it i.e.} following similar reasoning as that leading to the evaluation of (\ref{O2}),  
as will be discussed in detail in \cite{long}. 

The functions $a_i(\tau), b_i(\tau), E(\tau), S(\tau)$ may be computed in generality in terms of hypergeometric functions. The general expressions are too long to be reported here (they will be given in \cite{long}).
$E(\tau)$ diverges for $d \ge 3$, $S(\tau)$ for $d \ge 2$, and
$a_i(\tau)$ for $d \ge 3 + 2 i$, and $b_i(\tau)$ are finite, where we assume (as usual) $\Delta > d/2$.
When $d>6$ the theory is not renormalizable (as expected) as we need new counterterms of the schematic form $\partial^{2 n} \Phi^4$, with $n$ an integer.
(The renormalizability of the 4-point function at 1-loop order for $d=5, 6$ also holds in flat space \cite{long} and appears to be accidental).

The renormalized mass is given by
\begin{widetext}
\begin{align}
    m_R^2 = &m^2+\lim_{\tau \to 0}\left(\frac{\lambda}{2}\text{Con}\left[G_\tau(1)\right] 
    + \frac{\lambda^2}{4}\left( \text{Con}\left[\left(3\text{Div}\left[a_0(\tau)\right]+\frac{\left(2\Delta-\frac{d}{2}\right)}{\Delta}\text{Div}\left[a_1(\tau)\right]\right)G_\tau(1) \right] + 2 G_1 \text{Con}\left[G_\tau(1)\right]\right) \right.\nonumber\\
    &\left.-\frac{\lambda^2}{4}\text{Con}\left[\text{Con}\left[G_\tau(1)\right]E(\tau)\right]-\frac{\lambda^2}{6}\text{Con}\left[S(\tau)\right]-\frac{\lambda^2}{2}F_1\text{Con}[E(\tau)]\right)+\lambda F_1+\lambda^2F_2\, . 
\end{align}
\end{widetext}
This expression is scheme dependent  and one needs renormalization conditions to obtain physical results. 
Using $m_R^2 = \Delta_R (\Delta_R-d)$, one may work out the anomalous dimension $\gamma$, $\Delta_R= \Delta+\gamma$, perturbatively in $\lambda$. It remains to deal with the IR divergences. Apart from the integral in (\ref{tadpole}), there are also IR divergent integrals of the schematic form $\int G K$ and $\int \int K G K$, which are needed. The detailed evaluation of these integrals  will be presented in \cite{long}. The main result is that one may cancel all IR divergences by using $Z_\varphi = \varepsilon^{-\gamma}$ and the same boundary counterterm action $S_{\rm ct}$ as at tree-level but with $\Delta \to \Delta_R$.  

The  renormalization of $\Phi^4$ theory in AdS parallels that of $\Phi^4$ theory in flat space, which is discussed for example in chapter 4 of \cite{Ramond:1981pw} (actually, our notation for the scheme dependent functions matches that of \cite{Ramond:1981pw}). This is not unexpected as the short distance behavior of the 
theory should not depend on the large distance asymptotics. For example, the beta function for $\lambda$ matches exactly the beta function of $\Phi^4$ theory in flat space, as already noted in \cite{Bertan:2018afl}. One difference between the two cases is that here we need to renormalize the boundary source while in flat space we need wavefunction renormalization.

We are now in position to state the final results for the dual correlators.  The two-point function takes exactly the same form as the tree-level result\footnote{There is still a freedom of finite $\lambda$-depended rescaling of the source $\varphi$, which will change the normalization of the 2-point function. This freedom may also be thought as scheme-dependence.}  but with $\Delta \to \Delta_R$:
\begin{equation}
\langle \mathcal{O}(\vec{y}_1)\mathcal{O} (\vec{y}_2)\rangle= (2\Delta_R-d)c_{\Delta_R}
y_{12}^{-2\Delta_R}\, .
\end{equation}
The 4-point function, for the renormalizable cases, $d<7$,  is given by
\begin{widetext}
\begin{align}
    &\langle\mathcal{O}(\vec{y}_1)\mathcal{O}(\vec{y}_2)\mathcal{O}(\vec{y}_3)\mathcal{O}(\vec{y}_4)\rangle 
    = -(\lambda +\lambda^2G_1)c_{\Delta_R}^4D_{\Delta_R,\Delta_R,\Delta_R,\Delta_R}
    +\frac{\lambda^2}{2}c_{\Delta_R}^4\sum_{i=0}^\infty\left[\text{Con}\left[a_i(0)\right]D_{\Delta_R,\Delta_R,\Delta_R+i,\Delta_R+i}\ y_{34}^{2i} \right. \nonumber\\
&\left.\qquad +b_i(0)\frac{d}{d\alpha}\left(D_{\Delta_R,\Delta_R,\Delta_R+i+\alpha,\Delta_R+i+\alpha}\ y_{34}^{2i+2\alpha}\right)_{\alpha=0} + t{\text-}\  {\rm and}\ u\text{-channel} \right] = F(u,v) \prod_{i<j}
   y_{ij}^{-2 \Delta_R/3}\, ,
\end{align}
\end{widetext}
where 
$D_{\Delta_R,\Delta_R,\Delta_R,\Delta_R}$ is the tree-level contact diagram \cite{DHoker:1999kzh}. In the last equality 
we provide the answer in the form expected from conformal invariance
and the function of cross ratios $u, v$ is given by
\begin{widetext}
\begin{align}
    F(u,v) =& \frac{\pi^{d/2}}{2}\frac{c_{\Delta_R}^4}{\Gamma(\Delta_R)^2}(u v)^{\Delta_R/3} \left\{-(\lambda+\lambda^2G_1)\hat{H}_0+\frac{\lambda^2}{2}\sum_{i=0}^\infty\left[\text{Con}\left[a_i(0)\right]\hat{H}_i+b_i(0)\frac{d}{d\alpha}\left(\hat{H}_{i+\alpha}\right)_{\alpha=0} + 2\ {\rm perms}\right] \right\}\, ,
\end{align}
\end{widetext}
where $\hat{H}_i \equiv \frac{\Gamma\left(2\Delta_R-\frac{d}{2}+i\right)}{\Gamma(\Delta_R+i)^2}H(\Delta_R,\Delta_R,1-i,2\Delta_R;u,v)$ and the function
 $H(\alpha, \beta, \gamma, \delta; u, v)$ is given in (5.9) of  \cite{Dolan:2000uw} (see also \cite{Dolan:2000ut}) and  is related to the Appell $F_4$ hypergeometric function. The coefficients Con[$a_i(0)$] and $b_i(0)$ are explicitly computable; for example, when $d=3, \Delta=2$, $\text{Con}[a_i(0)]= 2 b_i(0) =-\delta_{i,0}/8 \pi^2$.

\paragraph{Conclusions:} We presented a systematic renormalization procedure for loop diagrams in AdS, and we illustrated the method using scalar $\Phi^4$ theory. Bulk renormalization is completely consistent with expectations based on the AdS/CFT duality and this provides further structural support for the duality.  It would be interesting to include graviton exchanges in the bulk and 
discuss tensorial correlators, as well as generalize the discussion to general $n$-point function, possibly to all loops. Finally, one should use the results obtained here in conjunction with recent results based on the conformal bootstrap.

\begin{acknowledgments}
{\em Acknowledgments.} 
M.B. would like to thank Glenn Barnich, Alberto Faraggi and Stefan Theisen for illuminative discussions. M.B., I.M. and E.B. were partially funded by Fondecyt Grant No. 1201145. I.M. also acknowledges financial support from Agencia Nacional de Investigación (ANID)-PCHA/Doctorado Nacional/2016-21160784 and would like to thank the Theoretical Physics \& Applied Mathematics group in Mathematical Sciences, University of Southampton for hospitality during part of this work. K.S. acknowledge support from the Science and Technology Facilities Council (Consolidated Grant “Exploring the Limits of the Standard Model and Beyond”). 
 
\end{acknowledgments}

\bibliography{references_master}

\end{document}